\documentclass[11pt,a4paper]{article}
\usepackage{mcite}
\usepackage{epsfig}

\epsfverbosetrue

\def\3{\ss}

\newcommand{\tev}{{\rm Te}\kern-1.pt{\rm V}}
\newcommand{\gev}{{\rm Ge}\kern-1.pt{\rm V}}
\newcommand{\mev}{{\rm Me}\kern-1.pt{\rm V}}
\newcommand{\kev}{{\rm Ke}\kern-1.pt{\rm V}}
\newcommand{\gevsq}{\mbox{$\mathrm{{\rm Ge}\kern-1.pt{\rm V}}^2$}}
\newcommand{\gevmsq}{\mbox{$\mathrm{{\rm Ge}\kern-1.pt{\rm V}}^{-2}$}}

\newcommand{\BR}        {\mbox{$\mathcal{B}$}}

\newcommand{\sla}[1]{/\!\!\!#1}

\begin{document}
\begin{titlepage}

\begin{center}
\begin{huge}
\bf \boldmath Prospects for the Observation of a Higgs Boson with $H\rightarrow\tau^+\tau^-\rightarrow l^+l^-\sla{p_t}$ Associated with One Jet at the LHC  \\
\end{huge}

\vspace{2.cm}

\Large B.~Mellado, W.~Quayle, Sau Lan Wu\\
\vspace{0.5cm}
{\Large\it Physics Department \\
University of Wisconsin - Madison \\
   Madison, Wisconsin 53706 USA }

\vspace{1.5cm}

\begin{abstract}
\noindent  The sensitivity of the LHC experiments to the Standard
Model Higgs using $H\rightarrow\tau\tau\rightarrow
l^+l^-\sla{p_t}$ associated with one high $P_T$ jet in the mass
range $110 <M_H<150\,\gev$/c$^2$ is investigated. A cut and Neural
Network based event selections are chosen to optimize the expected
signal significance with this decay mode. A signal significance of
about $6.6\,\sigma$ can be achieved for $M_H=120\,\gev$/c$^2$ with
$30\,$fb$^{-1}$ of integrated luminosity for one experiment only.
With this approach, experimental issues related to tagging forward
jets and to the application of a central jet veto are simplified
considerably.
\end{abstract}
\end{center}
\setcounter{page}{0}
\thispagestyle{empty}

\end{titlepage}

\newpage

\pagenumbering{arabic}

\section{Introduction}
\label{sec:introduction}

In the Standard Model (SM) of electro-weak and strong
interactions, there are four types of  gauge vector bosons (gluon,
photon, W and Z) and twelve types of fermions (six quarks and six
leptons)~\cite{np_22_579,prl_19_1264,sal_1968_bis,pr_2_1285}.
These particles have been observed experimentally. At present, all
the data obtained from the many experiments in particle physics
are in agreement with the Standard Model.  In the Standard Model,
there is one particle, the Higgs boson, that is responsible for
giving masses to all the
particles~\cite{prl_13_321,pl_12_132,prl_13_508,pr_145_1156,prl_13_585,pr_155_1554}.
In this sense, the Higgs particle occupies a unique position.

Prior to the end of the year 2000, the Higgs particle was not
observed experimentally.  After the center-of-mass energy at the
LEP accelerator of CERN reached $205\,\gev$ in 2000, excess
candidates began to show up in the Standard Model Higgs analysis
in the ALEPH experiment, consistent with a Higgs mass, $M_H$,
around $115\,\gev$/c$^2$~\cite{pl_495_1,rpp_65_465}.

One of the most exciting prospects for the LHC is confirming or
rejecting the first possible experimental evidence for the Higgs
particle at a mass around $115\,\gev$/c$^2$.

\begin{figure}[h]
{\centerline{\epsfig{figure=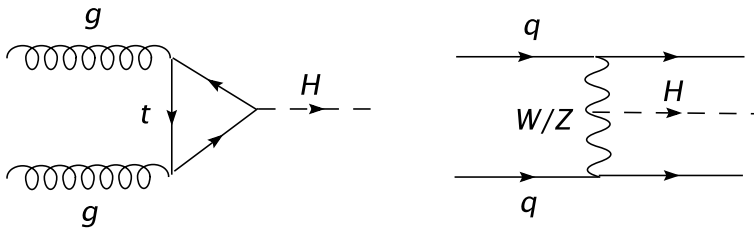,width=10cm}}}
\caption[]{Leading Order diagrams of the dominant processes
involving the production of Higgs at the LHC: Gluon-Gluon Fusion
(left) and Vector Boson Fusion (right).} \label{fig:higgs}
\end{figure}

The Standard Model Higgs will be produced at the LHC via several
mechanisms. The  Higgs will be produced predominantly via
gluon-gluon fusion~\cite{prl_40_11_692} (see left diagram in
Figure~\ref{fig:higgs}). For Higgs masses, such that
$M_H>100\,\gev$/c$^2$, the second dominant process is vector boson
fusion (VBF)~\cite{pl_136_196,pl_148_367} (see right diagram in
Figure~\ref{fig:higgs}).

Even for these dominant processes, the Higgs production
cross-section is small. Due to the large QCD backgrounds,
identifying the Higgs requires rejecting backgrounds by several
orders of magnitude. The identification of the Standard Model
Higgs boson with a mass around $115\,\gev$/c$^2$ is especially
challenging. Here, the most promising final states involve the
Higgs decays
$H\rightarrow\gamma\gamma$~\cite{CMS_ECAL_TDR,LHCC99-14,pl_431_410}
and $H\rightarrow\tau^+\tau^-$~\cite{pr_59_014037,pr_61_093005}.

In the case of Gluon-Gluon Fusion, shown in
Figure~\ref{fig:higgs}, the Higgs can be produced alone. However,
when one of the gluons or top quarks emits a gluon, the Higgs is
produced with the gluon, which is seen in the detector as a
hadronic jet. When the Higgs has a significant transverse momentum
the associated jet tends to be back-to-back with the Higgs in the
transverse plane (for reasons of transverse momentum balance). In
the case of Vector Boson Fusion, also shown in
Figure~\ref{fig:higgs}, the Higgs is produced with at least two
jets. In both cases, jets produced in association with the Higgs
are most useful in the identification of the Higgs, suppressing
significantly the QCD background.

Early  analyses performed at the parton level with the decays
$H\rightarrow\tau^+\tau^-\rightarrow l^+l^-\sla{p_t}$ associated
with two high transverse momentum, $P_T$, jets indicated that this
final state can be a powerful observation mode for a Higgs mass
around $115~\gev$/c$^2$~\cite{pr_59_014037,pr_61_093005}. The
ATLAS and CMS collaborations have performed feasibility studies
for these decay modes including more detailed detector description
and the implementation of initial state and final state parton
showers, hadronization and multiple interactions, which has
confirmed the strong potential of this final
state~\cite{CMS_NOTE_2003_033,SN-ATLAS-2003-024}. However, this
approach is intimately dependent upon the ability of tagging
forward jets and applying a stringent central jet veto. The
application of these experimental criteria will require a level of
knowledge of the detectors' response, which may not be attained
during the early stages of data taking at the LHC.

In this work we have explored the prospects of observing the
decays $H\rightarrow\tau^+\tau^-\rightarrow l^+l^-\sla{p_t}$
associated with one high $P_T$ jet in the final state (a different
approach was assessed within the context of Higgs searches at the
Tevatron~\cite{JHEP_0307_021}). We have demonstrated that this
final state enhances the potential for observing the Higgs at the
LHC. Additionally, with this approach, experimental issues related
to tagging forward jets and to the application of a central jet
veto are considerably simplified.

\section{Signal and Background Processes}
\label{sec:processes}

For simplicity, we have considered only the two main Higgs
production mechanisms at the LHC. Table~\ref{tab:sigcross} shows
the cross-sections for these processes estimated for a
center-of-mass of $14\,\tev$, as a function of the Higgs mass, in
the range $110<M_H<150\,\gev$/c$^2$. Shown are the Next-to-Leading
Order (NLO) cross-section for the gluon-gluon fusion mechanism
($\sigma_{gg}^{NLO}$), as calculated by
MC@NLO~\cite{JHEP_0206_029,JHEP_0308_007}, and the Leading-Order
(LO) cross-section for the VBF mechanism ($\sigma_{VBF}^{LO}$), as
calculated with PYTHIA6.2~\cite{cpc_82_74,cpc_135_238}. The proton
structure function parametrization CTEQ was used to evaluate the
proton-proton cross-sections~\cite{epj_12_375}.
Table~\ref{tab:sigcross} also displays the Higgs branching ratio
into a $\tau$ pair, $\BR(H\rightarrow\tau^+\tau^-)$, as calculated
with the HDECAY package~\cite{cpc_108_56}.

\begin{table}[ht]
\begin{center}
\begin{tabular}{||c|c|c|c||}
\hline
  $M_H (\gev$/c$^2$)   & $\sigma_{gg}^{NLO}$ (pb)  &  $\sigma_{VBF}^{LO}$ (pb)& $\BR(H\rightarrow\tau^+\tau^-)$\\ \hline

110 & 36.64 &  4.65  & 0.0765 \\

120 & 31.33 &  4.30  & 0.0685 \\

130 & 27.09 &  3.99  & 0.0539 \\

140 & 23.62 &  3.68  & 0.0355 \\

150 & 20.77 &  3.38  & 0.0183 \\

\hline

 \end{tabular}
 \caption{Cross-sections for $pp\rightarrow H+X$ for different Higgs masses. Values of the Next-to-Leading Order
cross-section for the gluon-gluon fusion mechanism,
$\sigma_{gg}^{NLO}$ and the Leading Order cross-section for the
VBF mechanism, $\sigma_{VBF}^{LO}$, are given. The Higgs branching
ratio into a $\tau$ pairs, $\BR(H\rightarrow\tau^+\tau^-)$, is
given in the last column.}
 \label{tab:sigcross}
\end{center}
\end{table}

The basic experimental signature of interest consists of:
\begin{itemize}
\item Two oppositely charged leptons (electron or muon) with large
transverse momentum. \item Large missing momentum, $\sla{p_t}$,
due to the presence of neutrinos in the final state. \item A large
transverse momentum jet (tag jet).
\end{itemize}

Two relevant SM background processes are considered, which
contribute to the final state specified above.
\begin{itemize}
\item $pp\rightarrow Z/\gamma^*+X$ production with
$Z/\gamma^*\rightarrow e^+e^-,\mu^+\mu^-,\tau^+\tau^-$ and
$\tau\rightarrow l\nu_{l}\nu_{\tau}$. After requiring the presence
of two charged leptons and large missing transverse momentum,
$pp\rightarrow Z/\gamma^*+X$ with
$Z/\gamma^*\rightarrow\tau^+\tau^-\rightarrow l^+l^-\sla{p_t}$
process is expected to be the dominant background. The inclusive
NLO cross-section of this process at the LHC yields $137\,$pb. A
strong suppression factor can be achieved with the application of
a dedicated event selection (see Sections~\ref{sec:eventselection}
and~\ref{sec:results}). \item $pp\rightarrow t\overline{t}+X$
production with $t\rightarrow Wb$, and $W\rightarrow
e\nu_{e},\mu\nu_{\mu},\tau\nu_{\tau}$ with $\tau\rightarrow
l\nu_{l}\nu_{\tau}$. The NLO cross-section at the LHC yields
$35.8\,$pb. This process is expected to display relatively large
associated jet multiplicity, hence, the rejection of this
background process is smaller with respect to that achieved for
the $pp\rightarrow Z\gamma^*+X$ process (see
Sections~\ref{sec:eventselection} and~\ref{sec:results}).
\end{itemize}

The contribution from events with at least one fake lepton in the
final state is neglected here, due to the large fake lepton
rejection expected to be attained with the CMS and ATLAS
detectors.

\section{MC Generation}
\label{sec:generation}

Events corresponding to the signal process with the VBF mechanism
have been generated with the LO matrix element based generator
provided by PYTHIA6.2. The rest of the processes under
consideration (Higgs via Gluon-Gluon Fusion and the background
processes pointed in Section~\ref{sec:processes}) have been
treated with MC@NLO, which implements NLO matrix elements
consistently matched with a parton shower generator. In order to
assess the impact on the production of $Z$ associated with hard
jets coming from diagrams with weak bosons in the internal lines,
the MadGraphII package was used~\cite{pc_81_357,hep-ph_0208156}.
This program incorporates tree-level matrix elements for the
process $pp\rightarrow Zjj$.

Higher order QCD corrections as well
as effects due to off-shell tops have not been addressed here.
These will be considered by the authors in future updates.

The impact of initial and final state QCD radiation,
hadronization, multiple interactions and underlying event were
simulated with PYTHIA6.2 and
HERWIG6.5~\cite{np_310_461,cpc_67_465,JHEP_0101_010}, depending on
the process. These MC programs were interfaced with the package
ATLFAST in order to simulate the response of the ATLAS
detector~\cite{ATLFAST}.



\section{Event Selection}
\label{sec:eventselection}

The following pre-selection cuts have been applied:
\begin{itemize}
\item [{\bf a.}] Require two high $P_T$ leptons in the central
region of the detector ($\left|\eta\right|<2.5$). Due to trigger
requirements, an event is accepted if at least one $\mu$ ($e$) has
$P_T>20\,\gev$ ($P_T>25\,\gev$) or if at least two leptons are
found with $P_T>10\,\gev$ for muons and $P_T>15\,\gev$ for
electrons. An average $90\%$ lepton identification efficiency is
assumed. At this stage, a central b-jet veto is applied in order
to suppress the contribution from $pp\rightarrow t\overline{t}+X$.
Events are vetoed if a jet consistent with a b-jet hypothesis is
found in the central region of the detector. An average $60\%$
b-jet tagging efficiency is assumed in this region of the detector
with rejections against c-jets and light jets of 10 and 100,
respectively. \item [{\bf b.}] To suppress the contribution from
$pp\rightarrow Z/\gamma^*+X$ with $Z/\gamma^*\rightarrow
e^+e^-,\mu^+\mu^-$ and to further suppress $pp\rightarrow
t\overline{t}+X$ production a cut on the invariant mass of the
leptons is applied, $M_{ll}<75\,\gev$. \item [{\bf c.}] In order
to reconstruct the mass of the Higgs candidate, it is assumed that
the decay products of the $\tau$'s are collinear to the $\tau$'s
themselves in the laboratory system~\cite{np_297_221} (usually
referred to as the collinear approximation). The variables
$x_{\tau1}$ and $x_{\tau2}$ are defined as the energy fraction of
the decaying $\tau$'s carried by the charged leptons. By using the
conservation of the transverse momentum, $\sum_{i=1,2}P_{T\tau
i}=\sum_{i=1,2}P_ {Tli}+\sla{p_{T}}$. The variables $x_{\tau1}$
and $x_{\tau2}$ are the solution of two linear equations. The mass
of the $\tau^+\tau^-$ pair, $M_{\tau\tau}$, can be computed as
$M_{\tau\tau}=M_{ll}/\sqrt{x_{\tau1}x_{\tau2}}$. It is required
that $0<x_{\tau1}, x_{\tau2}<1$.
 \item [{\bf d.}] The presence of at least one hadronic jet with
 $P_T>30\,\gev$ and $\left|\eta\right|<4.9$ is required.
\end{itemize}

\begin{figure}[t]
{\centerline{\epsfig{figure=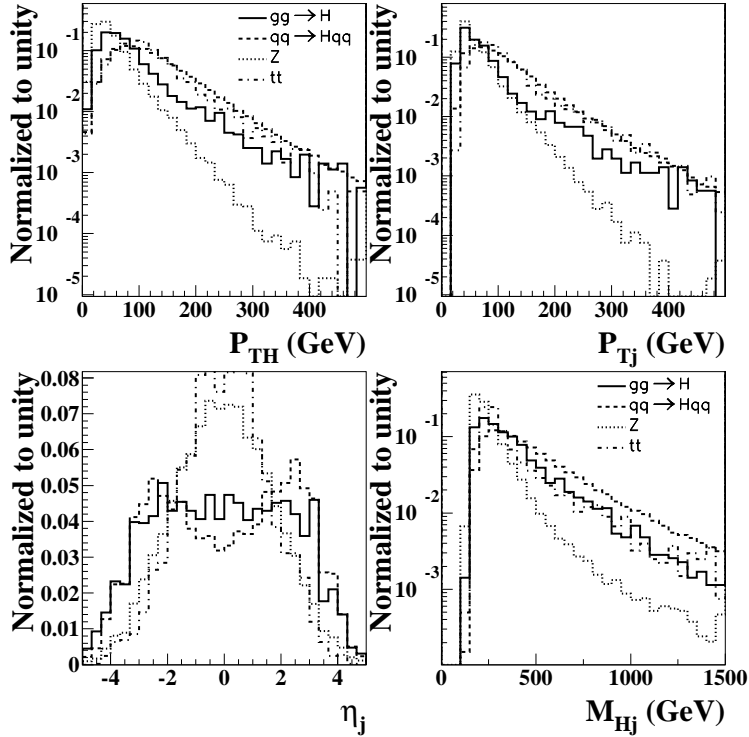,width=10cm}}}
\caption[]{Signal and background kinematic distributions after the
application of cuts {\bf a}-{\bf d} (see
Section~\ref{sec:eventselection}). An additional cut on the
invariant mass of the $\tau$ pair, $110<M_{\tau\tau}<150\,\gev$,
was applied. The upper left and right plots display the $P_T$ of
the Higgs candidate and the leading jet, respectively. The lower
left and right plots correspond to the pseudorapidity of the
leading jet and the invariant mass of the leading jet and the
Higgs candidate, respectively. The solid and dashed curves
correspond to Higgs production. The dotted-dashed and dotted
curves correspond to the two main backgrounds considered here (see
Section~\ref{sec:processes}). All curves were normalized to
unity.} \label{fig:tautauj_2}
\end{figure}

\begin{figure}[t]
{\centerline{\epsfig{figure=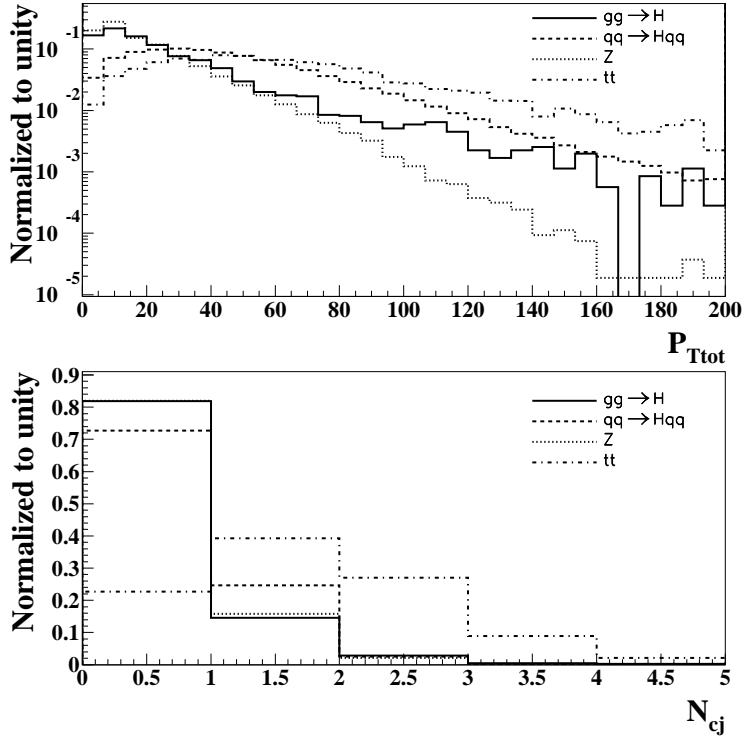,width=10cm}}}
\caption[]{Signal and background kinematic distributions after the
application of cuts {\bf a}-{\bf d} see
Section~\ref{sec:eventselection}). An additional cut on the
invariant mass of the $\tau$ pair, $110<M_{\tau\tau}<150\,\gev$,
was applied. The upper plot shows the transverse momentum of the
system made of the Higgs candidate and the leading jet. The lower
plot corresponds to the number of sub-leading central jets (see
Section~\ref{sec:eventselection}). The solid and dashed curves
correspond to Higgs production. The dotted-dashed and dotted
curves correspond to the two main backgrounds considered here (see
Section~\ref{sec:processes}). All curves were normalized to
unity.} \label{fig:tautauj_4}
\end{figure}

The discriminating power of various observables was examined after
the application of cuts {\bf a}-{\bf d}. This exercise was
performed for $M_H=120\,\gev$/c$^2$. In order to avoid biasses, an
additional cut on the invariant mass of the Higgs candidate is
applied, $110<M_{\tau\tau}<150\,\gev$.\footnote{This cut is
removed in the final evaluation of the signal significance. The
signal significance is calculated using a likelihood technique,
for which such a requirement is no longer necessary.} At this
stage, the background is largely dominated by $pp\rightarrow
Z/\gamma^*+X$ production (See Table~\ref{tab:cutflow} in
Section~\ref{sec:results}). Figures~\ref{fig:tautauj_2}
and~\ref{fig:tautauj_4} illustrate qualitative differences among
processes considered here. The solid and dashed curves correspond
to Higgs production via gluon-gluon fusion and VBF, respectively.
The dotted-dashed and dotted curves correspond to the two main
backgrounds considered here (see Section~\ref{sec:processes}). All
curves were normalized to unity in the ranges specified in the
plots.

The upper left and right plots in Figure~\ref{fig:tautauj_2}
display the $P_T$ of the Higgs candidate, $P_{TH}$, and the
leading jet, $P_{TJ}$, respectively. The distribution of $P_{TH}$
in $pp\rightarrow Z/\gamma^*+X$ is significantly softer with
respect to that of the other processes considered here. The lower
left plot in Figure~\ref{fig:tautauj_2} corresponds to the
pseudorapidity of the leading jet.  Leading jets in background
production tend to be significantly more central than in signal
production. Large $P_T$ $Z/\gamma^*$ production is dominated by
$qg\rightarrow qZ/\gamma^*$ processes, in which the quark in the
final state tends to be produced centrally. Higher order QCD
corrections in $pp\rightarrow t\overline{t}+X$ production also
favor central gluon radiation. On the other hand, large transverse
momentum production of Higgs via gluon-gluon fusion is mostly due
to initial state radiation off incoming gluons, which favors the
production of more forward jets. Finally, VBF represents the
t-channel exchange of two weak bosons by two quarks, which favors
the production of forward and very forward jets.

The lower plot in the right in Figure~\ref{fig:tautauj_2} displays
the invariant mass of the Higgs candidate and the leading jet in
the event, $M_{HJ}$. This variable is crucial to suppress
$pp\rightarrow Z/\gamma^*+X$ production. This variable, however,
fails to efficiently suppress $pp\rightarrow t\overline{t}+X$
production. Fortunately, the large jet multiplicity in
$pp\rightarrow t\overline{t}+X$ production not associated to the
actual top decay can be used to further suppress this background
process. The upper plot in Figure~\ref{fig:tautauj_4} shows the
$P_T$ of the system made up by the Higgs candidate and the leading
jet in the event.  The lower plot in Figure~\ref{fig:tautauj_4}
displays the number of sub-leading jets with $P_T>30\,\gev$ and
$\left|\eta\right|<2$, showing a qualitatively different behavior
in $pp\rightarrow t\overline{t}+X$ production. The application of
a veto on central sub-leading jets is very efficient in further
suppressing this background. It is worth noting that the
probability for $pp\rightarrow t\overline{t}+X$ events to survive
a veto on central jets does not depend strongly on the transverse
momentum cut applied on these jets.

The following additional cuts are added to the event selection:
\begin{itemize}
\item [{\bf e.}] Tagging jet is defined as the leading jet in the
event. It is required that the tagging jet not be very central,
$\left|\eta\right|>1$. Events are vetoed in which at least one
additional jet with $P_T>30\,\gev$ and $\left|\eta\right|<2$ is
reconstructed. \item [{\bf f.}] Tight cut on the Higgs candidate
transverse momentum, $P_{TH}>100\,\gev$. \item [{\bf g.}] Tight
cut on the invariant mass of the Higgs candidate and the tagging
jet, $M_{HJ}>700\,\gev$.
\end{itemize}

The event selection suggested in~\cite{pr_59_014037,pr_61_093005}
involves identifying events with two jets well separated in
pseudorapidity. By requiring this, the signal contribution is
dominated by Higgs production via VBF, in which, at least one of
the jets is a very forward one. Additionally, the authors
of~\cite{pr_59_014037,pr_61_093005} envision the application of a
stringent veto on events with central jets with $P_T>20\,\gev$.
The present event selection entails a number of crucial advantages
over the the event selection suggested
in~\cite{pr_59_014037,pr_61_093005}.
\begin{itemize} \item [{\bf 1.}] As illustrated in the lower left plot
in Figure~\ref{fig:tautauj_2}, the bulk of the events display a
tagging jet with $\left|\eta\right|<4$. Therefore, in this
approach there is no need to deal with very forward jets, which
are challenging to identify and calibrate at the LHC. Moreover,
the present event selection entails the presence of a very large
$P_T$ jets, which identification and calibration is significantly
easier. \item [{\bf 2.}] Because the suppression of the
$pp\rightarrow Z/\gamma^*+X$ relies on a tight cut on $M_{HJ}$ as
opposed to a stringent central jet veto, the requirements on
sub-leading central jets are substantially looser. This renders
the analysis more robust against the presence of additional jets
produced by the underlying event and pile-up. \item [{\bf 3.}]
Because of the requirement $P_{TH}>100\,\gev$ and the presence of
a very large $P_T$ jet, the invariant mass resolution improves by
$30\%$ with respect to that obtained with the event selection
suggested in~\cite{pr_59_014037,pr_61_093005}. This is mainly due
to the improvement in the resolution in the missing transverse
momentum.
\end{itemize}

\section{Results and Conclusions}
\label{sec:results}

\begin{figure}[t]
\vspace{-0.7cm}
{\centerline{\epsfig{figure=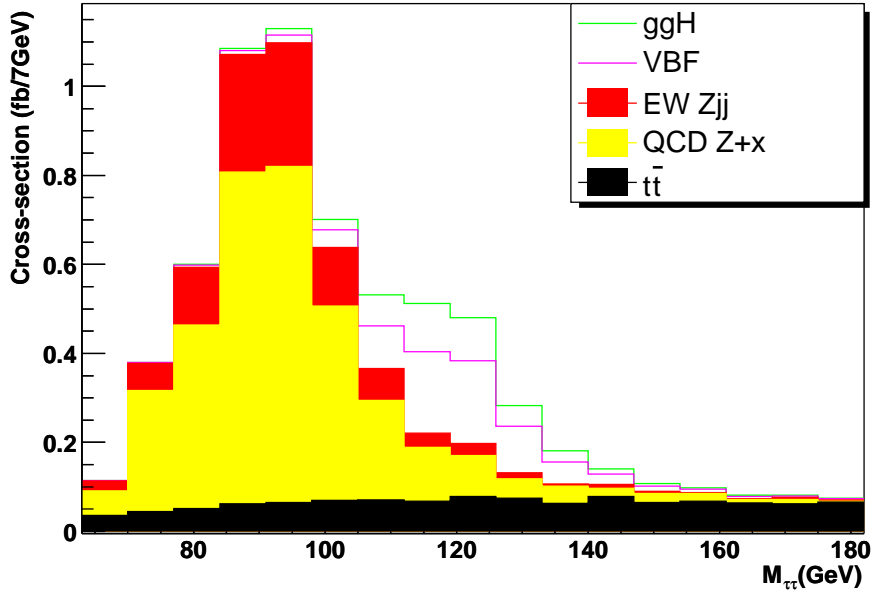,width=13cm}}}
\caption[]{Distribution of $M_{\tau\tau}$ (in fb/7\,\gev) after
the application of cuts {\bf a-g} given in
Section~\ref{sec:eventselection}. The solid line corresponds to
the total contribution of signal and background processes.
Histograms are staggered with the following order: $pp\rightarrow
t\overline{t}+X$, $pp\rightarrow Z/\gamma^*+X$ (the contribution
from Electro-weak $Zjj$ is given by the dark histogram), VBF $H$
and $gg\rightarrow H$.} \label{fig:mass}
\end{figure}

Table~\ref{tab:cutflow} gives the effective cross-sections after
the application of successive cuts in the event selection outlined
in Section~\ref{sec:eventselection}. Results are given in fb for
the signal processes with $M_H=120$ and the two major background
processes considered here. Results are also given after the
application of a mass window $110<M_{\tau\tau}<150\,\gev$. An
excellent signal-to-background ratio of about 1 can be achieved.
This is further illustrated in Figure~\ref{fig:mass}, which
displays the distribution of $M_{\tau\tau}$ (in fb/7\,\gev) after
the application of cuts {\bf a-g} given in
Section~\ref{sec:eventselection}. The solid line corresponds to
the total contribution of signal and background processes. The
dotted and dotted-dashed lines correspond to the total background
contribution and the contribution from $pp\rightarrow
t\overline{t}+X$ alone, respectively. The contribution from
$pp\rightarrow Z/\gamma^*+X$ with $Z/\gamma^*\rightarrow
e^+e^-,\mu^+\mu^-$ is negligible in the mass window. The
contribution of $pp\rightarrow Zjj$ from diagrams with weak bosons
in the internal lines contributed to about $20\%$ of the effective
cross-section for $pp\rightarrow Z/\gamma^*+X$ in the mass window.

\begin{table}[t]
\begin{center}
\begin{tabular}{||c|c|c|c|c|c|c||}
\hline
  Cut & $gg\rightarrow H$ & VBF $H$ &  \multicolumn{3} {c|} {$pp\rightarrow Z/\gamma^*+X$} & $pp\rightarrow t\overline{t}+X$ \\ \hline
{\bf a}  & 74.40 & 11.04 & 10.44$\times 10^{3}$ & 10.44$\times
10^{5}$ & 43.22 & 5.60$\times 10^{3}$ \\ \hline


{\bf b} & 67.20 & 10.22 & 10.32$\times 10^{3}$ & 10.39$\times
10^{4}$ & 41.84 & 1760 \\ \hline


{\bf c} & 47.3 & 8.91 & 5690 & 2.34$\times 10^{4}$ & 32.13 & 350 \\
\hline


{\bf d} & 26.51 & 8.57 & 1870 & 2440 & 31.40 & 347 \\
\hline


{\bf e} & 16.73 & 4.93 & 1030 & 1370 & 12.21 & 46.43 \\ \hline


{\bf f} & 1.72 & 2.05 & 81.6 & 25.2 & 3.38 & 16.66 \\ \hline


{\bf g} & 0.43 & 0.76 & 3.22 & 0.60 & 1.11 & 5.48 \\

\hline\hline


& 0.32 & 0.59 & 0.38 & 0 & 0.11 & 0.41 \\

\hline

 \end{tabular}
 \caption{Effective cross-sections (in fb) for signal ($M_H=120\,\gev$/c$^2$) and background processes after the applications of cuts {\bf a-g} specified in Section~\ref{sec:eventselection}. The effective cross-sections in a mass window $110<M_{\tau\tau}<150\,\gev$ are reported in the last row. The three columns for the
 process $pp\rightarrow Z/\gamma^*+X$, correspond to the contribution from QCD $Z/\gamma^*\rightarrow\tau^+\tau^-$, $Z/\gamma^*\rightarrow e^+e^-,\mu^+\mu^-$ and Electro-Weak $pp\rightarrow Zjj$ production, respectively.}
 \label{tab:cutflow}
\end{center}
\end{table}

Figure~\ref{fig:mass} illustrates the different shapes of the
$M_{\tau\tau}$ distributions for the two major backgrounds. While
the contribution from $pp\rightarrow t\overline{t}+X$ is
relatively flat, the contribution from $pp\rightarrow
Z/\gamma^*+X$ displays a steep behavior, which is a combination of
resolution effects and the presence of large mass $Z/\gamma^*$
production.

It is evident from Figure~\ref{fig:mass} that a good control of
the shape of the $M_{\tau\tau}$ distribution is crucial for
establishing a compelling deviation from a purely background
hypothesis. For the purpose of detailed studies of background
estimation, two different event selections with little overlap
were defined to address the two main backgrounds. The background
normalization and shape of the $M_{\tau\tau}$ can be studied in
these control samples and then extrapolated to the region of the
phase space under study, as suggested in~\cite{hep-ph_04_04045}.
It has been demonstrated that by means of an analysis based on LO
Matrix Elements the normalization of top backgrounds in
$H\rightarrow W^+W^-\rightarrow l^+l^-\sla{p_t}$ searches at the
LHC can be determined with an accuracy of better than
$10\%$~\cite{hep-ph_04_04045}.

\begin{table}[ht]
\begin{center}
\begin{tabular}{||c||c|c|c|c||}
\hline
  Control Sample & $gg\rightarrow H$ & VBF $H$ &  $pp\rightarrow Z/\gamma^*+X$ & $pp\rightarrow t\overline{t}+X$ \\ \hline
$pp\rightarrow Z/\gamma^*+X$ & 0.41 & 0.29 & 65.2 & 5.38 \\

$pp\rightarrow t\overline{t}+X$ &  0.21 & 0.03 & 0.51 & 40.5 \\
\hline

 \end{tabular}
 \caption{Effective cross-sections (in fb) for signal ($M_H=120\,\gev$/c$^2$) and background processes after the applications of cuts for two control samples to study major backgrounds.}
 \label{tab:control}
\end{center}
\end{table}

The control sample to study $pp\rightarrow Z/\gamma^*+X$ includes
all cuts presented in Section~\ref{sec:eventselection} except for
requiring that the tagging jet be very central,
$\left|\eta\right|<1$, and a change in cut {\bf g}, where
$250<M_{HJ}<400\,\gev$ is used instead. As for the control sample
to study $pp\rightarrow t\overline{t}+X$, the requirement of a
b-jet veto is removed from cut {\bf a}, in cut {\bf e} the tagging
jet is requited to be central, $\left|\eta\right|<2.5$, and no
veto on additional jets is applied. In addition, the transverse
momentum of the system made by the Higgs candidate and the tagging
jet is required to be larger than $100\,\gev$.
Table~\ref{tab:control} displays the contribution from signal and
background processes after the application of these two event
selections. Both event selections provide relatively clean
environments to study the corresponding backgrounds. With
30\,fb$^{-1}$ of integrated luminosity the background
normalization will be understood to better than $10\%$.

The expected signal significance was calculated using a likelihood
technique~\cite{ATL-PHYS-2003-008,physics_03_12050}.
Table~\ref{tab:results} shows the expected signal significance as
a function of the Higgs mass with 30\,fb$^{-1}$ of integrated
luminosity (for one experiment only). A $10\%$ systematic error on
the background estimation has been assumed.

Table~\ref{tab:results} also reports the results obtained with a
multivariate analysis performed with the help of Neural Network
(NN) algorithms. For this purpose, NN's have been trained using
the following discriminating variables: pseudorapidity of the
tagging jet~\footnote{The requirement that $\left|\eta\right|<1$
placed in cut {\bf e} (see Section~\ref{sec:eventselection}) was
removed for the NN training.}, $M_{HJ}$ and $P_{TH}$. A net
improvement in the signal significance of about $25\%$ over the
classical cut analysis can be achieved.

\begin{table}[ht]
\begin{center}
\begin{tabular}{||c||c|c|c|c|c||}
\hline
  Higgs Mass $(\gev$/c$^2$) & 110 & 120 & 130 & 140 & 150  \\ \hline
Signal Significance for cut analysis ($\sigma$) & 4.3 & 5.0 & 4.8 & 3.6 & 2.1 \\
Signal Significance for NN analysis ($\sigma$) & 5.5 & 6.6 & 6.3 & 4.8 & 2.8 \\
\hline
 \end{tabular}
 \caption{Expected signal significance for $H\rightarrow\tau^+\tau^-\rightarrow l^+l^-\sla{p_t}$ associated with one high $P_T$ jet as a function of the Higgs mass with 30\,fb$^{-1}$ of integrated luminosity (one experiment only). A $10\%$ systematic error on the background estimation has been assumed. Results are given for the cut and NN based analyses.}
 \label{tab:results}
\end{center}
\end{table}

The feasibility of searches for Minimal Super-Symmetric Higgs at
the LHC with this final state needs to be investigated. It is
worth noting that the approach presented in this work can be
applied to other Higgs decays~\cite{WWJ}. Generally speaking, this
approach can be applied in searches of particles produced via the
t-channel exchange of weak bosons.

\section{Acknowledgments}
\label{sec:acknowledgements} The authors are most grateful to
S.~Frixione, T.~Han, K.~Jakobs, N.~Kauer, T.~Plehn, D.~Rainwater,
T.T.~Wu and D.~Zeppenfeld for comments and suggestions. The
authors want to thank for ATLAS Collaboration for the support and
encouragement. This work was supported in part by the United
States Department of Energy through Grant No. DE-FG0295-ER40896.

\bibliographystyle{zeusstylem}
\bibliography{vbf,mycites}

\end{document}